**Formation of Graphene on SiC($000\bar{1}$) Surfaces in Disilane and Neon Environments**

Guowei He, N. Srivastava, R. M. Feenstra[*]
Dept. Physics, Carnegie Mellon University, Pittsburgh, PA 15213

**Abstract**
The formation of graphene on the SiC($000\bar{1}$) surface (the *C-face* of the {0001} surfaces) has been studied, utilizing both disilane and neon environments. In both cases, the interface between the graphene and the SiC is found to be different than for graphene formation in vacuum. A complex low-energy electron diffraction pattern with √43×√43-R±7.6° symmetry is found to form at the interface. An interface layer consisting essentially of graphene is observed, and it is argued that the manner in which this layer covalently bonds to the underlying SiC produces the √43×√43-R±7.6° structure [i.e. analogous to the 6√3×6√3-R30° "buffer layer" that forms on the SiC(0001) surface (the *Si-face*)]. Oxidation of the surface is found to modify (eliminate) the √43×√43-R±7.6° structure, which is interpreted in the same manner as the known "decoupling" that occurs for the Si-face buffer layer.

**I. Introduction**
Graphene formation on the {0001} surfaces of SiC has been intensively studied for the past 5 years, since that graphene can be formed with large areas (and on semi-insulating SiC) suitable for device and circuit fabrication.[1] By heating the SiC to temperatures about 1200 °C, Si atoms preferentially sublimate from the surface, leaving behind excess C atoms that self assemble into graphene. Of the two {0001} surfaces, much more is presently known concerning graphene formed on the (0001) surface (the *Si-face*) than for the ($000\bar{1}$) surface (the *C-face*). A number of groups have succeeded in forming single-monolayer (ML) graphene on the Si-face, with good reproducibility between the groups. The main method used is to form the graphene by heating the SiC in an argon environment, with the role of the argon being to reduce the sublimation rate of Si from the surface and hence allow increased heating temperatures (for a given graphene thickness, i.e. 1 ML).[2,3,4]

In contrast, for the C-face, it is difficult to form single-ML graphene in an argon environment. Rather, a number of studies reveal islands of graphene on the surface.[5,6,7] We have argued previously that those results are affected by unintentional oxidation of the C-face SiC surface, making it resistant to graphitization.[7] (The C-face appears to be more susceptible to oxidation than the Si-face, which we interpret in terms of the difference in structures for the two surfaces: the terminating layer on the Si-face is a 6√3×6√3-R30° structure,[8,9] which apparently has good stability and quite low energy, whereas the C-face is terminated in many cases with a 3×3 or 2×2 structure,[10] which appears to be less stable). Thus, to restrict the sublimation rate of the Si from the C-face, while simultaneously minimizing any unintentional oxidation, it is necessary to perform the heating in a cleaner environment. We accomplish that in the present work utilizing either disilane[11] at a pressure of ≈$10^{-4}$ Torr, or cryogenically-purified neon at 1 atm pressure.

---

[*] feenstra@cmu.edu

In these environments, we find that a new interface structure forms between the graphene and the C-face SiC, one that has √43×√43-R±7.6° symmetry (denoted by √43 for short). This structure was reported in our previous publication,[12] where we argued that the first C-rich layer that forms on the surface has this √43 structure and forms a graphene "buffer layer" that is covalently-bonded to the underlying SiC.[8] In the present work, we provide data from several samples that were not previously described (or only described briefly). The first of these is a C-face surface prepared in disilane, for which we employed rapid transfer of the sample between our graphene preparation system and the characterization system, thus enabling study the graphene buffer layer before any oxidation occurs underneath that layer (i.e. decoupling it from the SiC). Selected-area diffraction of the buffer layer reveals spots at identical locations to those of graphene, thus proving that the buffer layer does indeed have a structure very close to that of graphene. The other two samples were prepared in a neon environment. One of these is a Si-face sample, for which we find very similar results as for Si-face preparation in disilane or argon. The other sample is a C-face surface, prepared in neon, for which we find results similar to C-face preparation in disilane but quite different from that obtained in argon, i.e. due to the unintentional oxidation that occurs during the C-face preparation in argon.

This paper is organized as follows. In Section II we present experimental details. Section III describes our results, first for the disilane environment and then for the neon environment. A discussion of our results, and comparison of them with data of other groups, is presented in Section IV. The paper is summarized in Section V.

**II. Experimental**

Experiments were performed on nominally on-axis, n-type 6H-SiC or semi-insulating 4H-SiC wafers purchased from Cree Corp., with no apparent differences between results for the two types of wafers. The wafers were cut into 1×1 cm$^2$ samples. To remove polishing damage, the samples were heated in either 1 atm of hydrogen at 1600 °C for 3 min or 5×$10^{-5}$ Torr of disilane at 850 °C for 5 min. In the same chamber, graphene was formed by heating either in 5×$10^{-5}$ Torr of disilane or 1 atm of neon. Growth details for all samples are presented in Table I. For the preparation in neon, the neon was cryogenically purified by flowing liquid $N_2$ between the walls of the double-walled chamber, thus inhibiting unintentional oxidation of the C-face SiC (which was a problem in prior work using argon[7]). Characterization by low-energy electron diffraction (LEED) was performed *in situ* in a connected ultrahigh-vacuum chamber. After transferring the samples through air, further characterization was performed using an Elmitec III low-energy electron microscope (LEEM). Atomic force microscopy (AFM) was performed in air, using a Digital Instruments Nanoscope III in tapping mode.

**III. Results**
**A. Disilane Environment – C-face**
**1. Graphene on bare (unoxidized) SiC**
Our results for graphene formation in disilane for the C-face (and the Si-face) have been the subject of several prior works.[12,13,14] In this Section, we discuss results obtained from a new sample that was transferred from our graphene preparation system to the LEEM in a relatively short time, ≈10 minutes, as opposed to the several hours or more used for prior samples. Figure 1 shows results from this sample, denoted D1 in Table I, prepared by heating in disilane ($5\times10^{-5}$ Torr) at 1220°C for 10 min. The LEEM image at 3.8 eV shown in Fig. 1(a) consists predominantly of two types of areas, one with bright and the other with dark contrast. The LEEM image displays varying contrast due to different numbers of graphene layers on the surface. Measurements of the reflected intensity of the electrons as a function of their energy from locations marked in Fig. 1(a) are shown in Fig. 1(b). As demonstrated by Hibino et al., the number of minima in the reflectance curves corresponds to the number of graphene layers on the surface.[15] The reflectivity characteristics from the dark region, curves C and D of Fig. 1(b), reveal a single minimum near 3.7 eV, as is typical for single-ML graphene.[15] The reflectivity from the bright region, curves A and B of Fig. 1(b), show behavior that we have never observed previously on any C-face or Si-face sample (it has, however, been observed in a single sample studied by other workers,[16] as further discussed in Section IV). Small areas of this surface consist of multilayer graphene, as in curves E and F, which reveal 2 and 4 ML of graphene, respectively.

Additional information is contained in the selected-area diffraction (μ-LEED) results of Figs. 1(c) and 1(d). These patterns were acquired with a 5 μm aperture, at locations centered around the points A and C in Fig. 1(a). The size of that aperture is slightly larger than the areas of bright or dark contrast, respectively, surrounding those points, but data with a 2 μm aperture reveal the same diffraction spots (albeit with worse signal-to-noise) at these locations and at many other locations studied on the surface. At all locations, the patterns reveal spots with wavevector magnitude precisely equal to that of graphene.

Based on the μ-LEED results, we can be certain that surface structure leading to the reflectivity curves A and B has the structure of graphene However, we also know that it is the bottommost graphene layer, i.e. directly in contact with SiC, since extensive studies both on sample D1 and on many other samples (some with average graphene thickness less than that of sample D1)[12,13,14,17] reveal only thicker graphene or no graphene at all. Being in contact with the SiC, the electronic properties of this layer may differ from that of ideal graphene, and we hence refer to it as a "buffer layer", in analogy to the $6\sqrt{3}$ buffer layer on the Si-face.[8,9]

Immediately after sample D1 was produced, its LEED pattern acquired *in situ* at 100 eV revealed a $\sqrt{43}\times\sqrt{43}$-R±7.6° pattern, shown in Fig. 2(a). We interpret this pattern as indicating some distortion of the buffer layer and/or underlying SiC layer, due to bonding between the buffer layer and the SiC, again in analogy with the $6\sqrt{3}$ structure of the Si-face. It should be noted that the $\sqrt{43}$ spots that appears in LEED patterns acquired at 100 eV from these unoxidized samples are *not* seen in the lower-energy LEED patterns of Fig. 1(c) and 1(d) acquired in the LEEM. We attribute this discrepancy to a reduced sensitivity of those diffraction intensities in the LEEM measurements, as further discussed in Section IV.

**2. Graphene on oxidized SiC**

After the studies described in the previous Section, sample D1 was removed from the LEEM instrument and oxidized by exposing it to air for several days. This oxidation caused the $\sqrt{43}$ spots to disappear, as shown in Fig. 2(b) (for other samples, we sometimes also observe the formation of √3×√3-R30° spots for the oxidized surface,[12] but not for this particular sample, as further discussed in Section IV). LEEM results from the air-exposed surface are shown in Fig. 3. These LEEM images were not acquired from the same surface location as in Fig. 1 (due to difficulty in finding the same location), but nevertheless the surface of Fig. 3 was covered predominantly with areas of two different contrast levels, just as for Fig. 1, and we can confidently assign the two types of areas with the corresponding ones in Fig. 1.

The areas of bright and dark contrast can be seen in the LEEM image at 3.8 eV displayed in Fig. 3(a), although the dark areas now appear with two slightly different contrasts. Reflectivity curves from these dark regions, curves C and D of Fig. 3(b), reveal single-ML behavior (curve D) for the darkest contrast and single-ML plus an additional minimum at 6.9 eV (curve C) for the slightly lighter contrast areas. This behavior is exactly the same as in our previous work,[12] with the minimum at 6.9 eV interpreted as forming because of "decoupling" of the buffer layer that is below the single-ML graphene (i.e. release of the covalent bonds between the buffer layer and the underlying SiC, as in Refs. [18] and [19], due to oxidation of the SiC in the present case).

The reflectivity curves from the lighter-contrast areas in Fig. 3(a), curves A and B of Fig. 3(b), reveal the same behavior as we previously associated with the oxidized buffer layer on the C-face.[12] A broad maximum is seen over 2 – 6 eV, along with a minimum near 6.6 eV. We associate the minima at 6.6 eV for curves A and B and the one at 6.9 eV for curve C with the same origin, i.e. some feature arising from the oxidation of the buffer layer, which persists even with one (or more) graphene layer on top of the buffer layer. It is notable that for graphene layers on top of a decoupled *Si-face* buffer layer, a minimum in reflectivity is also observed near 7 eV.[20] This minimum has been suggested to arise from the detailed structure of the interface,[20] an interpretation that we believe likely holds for our C-face case as well.

**B. Neon Environment – Si-face**

As described in Sections I and II, the use of cryogenically-purified neon provides an effective means of restricting the Si sublimation rate from the surface without producing the unintentional oxidation that we find with the use of argon. We have prepared graphene on both the Si-face and the C-face in the neon environment, with results for the former case discussed in this Section. Figure 4 shows results for sample N1, prepared by heating to 1630 °C in the 1-atm neon environment. (These results were obtained after the sample was exposed to air for many days between preparation and LEEM study, although for Si-face graphene such air exposure does not lead to any noticeable change in the surface or interface structure). The surface morphology as shown in the AFM data of Fig. 4(a) consists of step bunches distributed over the surface, very similar to what occurs for preparation in argon.[7] The graphene thickness is found to be mainly 1 ML and 2 ML for these preparation conditions, along with a few 3 ML areas (likely near the step edges), again similar to what occurs for argon. LEED patterns at 100 eV of this surface (not shown) are nearly identical to those seen for argon-prepared graphene, as in Fig. 3 of Ref. [13], with intense 6√3 satellite peaks surrounding the main SiC and graphene peaks.

**C. Neon Environment – C-face**

Turning now to graphene on the C-face prepared in neon, sample N2, Fig. 5 shows an AFM image of the surface and Fig. 6 displays LEED results. This is the same sample discussed in Ref. [12], which displayed LEEM images of the surface. The graphene coverage on this surface consists mainly of areas of exposed buffer layer and 1-ML graphene on the buffer, although areas of multilayer graphene as well as some SiC without any buffer layer were also seen. In the AFM we see the characteristic ridges (white lines) on the surface, typical for graphene that has reasonably large grain size on the surface.[21] These ridges appear somewhat broken up in some locations, but that effect follows the scan direction of the AFM and is surely an artifact of the scanning.

In LEED, immediately after the graphene formation, the characteristic √43 LEED pattern is observed, as shown in Fig. 6(a). After oxidation of the surface (accomplished by exposing the surface to $1 \times 10^7$ L of oxygen with the sample at about 200 °C, after which it was briefly heated to 1000 °C), the √43 pattern disappears and some √3 spots appear, as seen in Fig. 6(b). This behavior is the same as typically found for C-face surfaces prepared under sufficiently high disilane pressure.[12] µ-LEED measurements of the oxidized surface, Figs. 6(c) and (d), reveal spots with wavevector equal to that of graphene. A significant result here is that, on the buffer layer, the µ-LEED acquired with 5 µm aperture reveals, in most cases, only a single sixfold arrangements of spots, as seen in Fig. 6(c). This result is in contrast with the disilane-prepared buffer layer, Fig. 2, in which multiple orientations of the sixfold pattern are observed. Thus, the crystallographic grain size for the neon-prepared graphene is found to be significantly larger than 5 µm. This is the best structural result that we have found to date in any of our graphene prepared on the C-face SiC.

It is also important to note that the orientation of the sixfold arrangement of spots in Fig. 6(c) is, judging from the 100-eV wide-area LEED result of Fig. 6(b), aligned at 30° relative to the primary (1,0) SiC spots (we further discuss this result in the following Section and compare it to that recently obtained by other workers[22]). This result is in contrast to the rotational orientation of the graphene on top of the buffer layer, Fig. 6(d), for which a range of orientation angles is found, and with this range being centered around the same azimuth as the SiC spots (i.e. the usual orientation for C-face graphene[1]).

**IV. Discussion**

The results presented in this paper for graphene prepared on the C-face (and the Si-face) are in good agreement with those presented in our previous works.[12,13,14] The results shown in Section III(A) for sample D1 go beyond those previous result by revealing results for the buffer layer structure *prior to* any significant oxidation of the surface. The reflectivity curves of that pristine buffer layer [curves A and B of Fig. 1(b)] appear quite different than after oxidation has occurred [curves A and B of Fig. 3(b)]. The general effect of the oxidation is seen to be the formation of a prominent minimum in the reflectivity near 6.6 eV. This minimum also occurs for graphene on top of the buffer layer, albeit shifted upwards by about 0.3 eV, as seen for 1-ML-thick graphene on the buffer layer [curve C of Fig. 3(c)].

For the pristine buffer layer, we find that µ-LEED reveals diffraction spots with the *same* wavevector magnitude as expected for perfect graphene. Thus, we conclude that the buffer layer is essentially a graphene layer, but one that is most likely covalently bonded to the underlying SiC. This situation is thus completely analogous to the 6√3 buffer that forms on the Si-face.[8,9,18,19] (It should be noted, however, that the angle-resolved photoemission experiments for the Si-face that directly reveal the bonding and subsequent decoupling of the buffer layer from the Si-face SiC, Ref. [18], are not yet available for the C-face; such data would provide more complete evidence for our interpretation of our C-face results).

When oxidation of our C-face samples occurs, μ-LEED results from the buffer layer are unchanged, as expected since that layer is essentially graphene. As noted in Section III(A), those μ-LEED results are acquired at energies of around 44 eV or below, since at higher energies the diffraction spots observed in the LEEM broaden considerably and lose intensity. For the wide-area LEED patterns acquired with a conventional LEED apparatus, and displayed in this work at 100 eV, a change *is* observed before and after oxidation of the sample, namely, the √43 spots are present in the former case but absent in the latter. We attribute this difference between the diffraction results of the LEEM compared to the conventional apparatus simply to the energy-dependence of the diffraction intensities (and also considering instrumental effects in the LEEM at the higher energies). Indeed, using the conventional LEED apparatus at lower energies, we find that the √43 spot intensities diminish as the energy is reduced from 100 eV, being very weak at energies below 60 eV. The disappearance of the √43 spots upon oxidation is interpreted, as already mentioned above, in terms of a release of the covalent bonds between the buffer layer and the underlying SiC.

For most C-face samples prepared in disilane we find that, upon oxidation, distinct √3 and 2√3 spots [i.e. (1/3,1/3) and (2/3,2/3)] appear in the diffraction.[12] In some cases [such as Fig. 2(b)] we do not observe those spots, although the reflectivity characteristics still indicate a decoupling of the buffer layer. We therefore believe in those cases that oxidation has occurred, but the oxidized layer has not ordered sufficiently to form the √3×√3-R30° structure. We have previously interpreted the √3 and 2√3 spots in terms of the formation of a silicate layer below the buffer layer, and we feel that this interpretation is still the most likely. (On samples with partial graphene coverage, such as the one described in Figs. 1 and 3 of Ref. [12], it is possible that some areas of *bare* SiC, i.e. not covered even by the buffer layer, may have existed, and those areas would certainly contribute strongly to the observed √3 and 2√3 spots, suggesting that perhaps the silicate structure is *not* present below the buffer layer itself. However, even on those samples, the intensity vs. energy characteristics of the √3 and 2√3 spots are slightly different than for an oxidized bare SiC surface without graphene,[7] which we take as qualitative evidence that the silicate layer does indeed exist beneath the buffer layer. Quantification of these characteristics is in progress). We note that a significant variation from sample to sample is found in the efficacy with which the √3×√3-R30° pattern forms. Samples with more complete graphene coverage are found to be more resistant to oxidation, a fact that we tentatively interpret in terms of a reduced diffusion probability of the oxygen through the complete graphene layer.

For the Si-face, results for sample N1 prepared in the neon environment are in good agreement with the previous reports employing preparation in argon.[3,4,7] The use of neon as compared with argon does not appear to significantly affect the situation for the Si-face. But for the C-face, our results using neon are in contrast to our earlier work in argon: we find graphene with relatively uniform coverage over the surface using neon, whereas severe islanding of the graphene occurred with argon.[7] We interpret this difference in terms of a cleaner environment for the cryogenically purified neon; our previous C-face work in argon displayed ample evidence for the existence of background oxygen during the graphene formation,[13] and apparently the amount of that oxygen is substantially lower in our neon environment.

It should be noted that, whereas several other groups have presented results for C-face graphene formation that we believe also display effects of unintentional surface oxidation,[5,6] one group has in fact shown results that, for a particular sample,[16] are very similar to what we obtain for our work in disilane.[12] Johansson et al. obtain, for C-face graphene prepared in argon, a reflectivity curve nearly identical to our pristine buffer layer curves A and B in Fig. 1(b).[16] But they present a much different interpretation of their results than that given here. Those workers interpret the reflectivity curves as arising from areas containing a mixture of graphene and a silicate (Si-rich oxide) structure,[16] but we believe that our identification of the existence of the buffer layer is clear from our data (Johansson et al. observed just a small area of this structure). Apparently, the argon environment of Johansson et al. is somewhat cleaner than our own, something that we find entirely possible (i.e. with greater purity of the gas and/or or better cleanliness of the gas lines and chamber).

Regarding cleanliness of the environment during graphitization, we believe that the use of disilane (or some other Si-containing environment) may offer a significant advantage. As discussed in Ref. [13], we have for a period of time used our graphitization system under conditions of reduced cleanliness due to the failure of a pump. In this condition, some of our processes such an H-etching did not work well, producing surfaces that exhibited some oxide. However, the surface cleaning in disilane and the subsequent graphitization in disilane worked normally, with no trace of surface oxide. We therefore suggest that the disilane (or Si) itself may act to scrub oxygen from the system, e.g. by the formation of volatile $SiO_x$ species. This type of reaction may be significant in systems with only moderate or low vacuum environment, such as the confined-controlled sublimation (CCS) process employed by de Heer and co-workers.[22]

In the CCS process, a small SiC sample is graphitized while contained within a graphite container that is nearly closed except for a small hole in its cap.[22] The silicon pressure in the container during the heating is expected to be substantial, estimated at $\approx 10^{-3}$ Torr.[22] The resulting quality of the graphene layer appears to be quite good, both structurally and electronically.[22] However, to scale up the process to larger wafer sizes (and to ensure reproducibility between graphitization systems), one would like to perform this process in an open vacuum system with known partial pressures of the various gaseous constituents. It is this goal that we have pursued in the present work and our prior work.[12] As mentioned in the previous paragraph, the chemical role of Si on the environment may be a significant one not only for its impact in reducing the Si sublimation rate, but also for its possible effect in maintaining an appropriate (i.e. reduced oxygen content) background gas.

The group of de Heer and co-workers have reported, for their best C-face single-ML graphene, a diffraction pattern consisting of sharp graphene spots located at positions rotated by 30° relative to the principle (1,0) SiC spots.[22] With that work as motivation, we have examined our LEED patterns for the same sort of result. In many cases, our graphene displays streaks, centered at typically ±7° from the azimuth of the principle SiC spots.[7,12,13] This spread in angles is well known for graphene on the C-face (although sometimes the angles are smaller), and has been associated with rotational disorder in the graphene layers.[1] However, for sample N2 shown in Fig. 6, after oxidation it displays reasonably sharp spots along a 30° azimuth relative to the SiC spots. μ-LEED of this surface also displayed predominantly only six spots in each pattern (i.e. no multiple domains in the 5 μm viewing area), as discussed in Section III(C), and the rotational alignment of those spots tends to be all the same. Thus, for that sample, it appears that the decoupled buffer layer (i.e. which itself is the "zeroth" graphene layer) is rotationally aligned at 30° relative to the SiC, in agreement with the result of Ref. [22].

## V. Conclusions

By forming graphene on the SiC($000\bar{1}$) surface in either disilane or purified neon environments, a new interface structure with √43×√43-R±7.6° symmetry is found to form between graphene and the underlying SiC substrate. The first C-rich layer above the SiC is interpreted as a "buffer layer", in analogy to what occurs on the SiC(0001) surface.[8,9] Before oxidation of the surface, this buffer layer is found to have a structure close to that of graphene. The √43×√43-R±7.6° symmetry is interpreted as arising from bonding to the underlying SiC substrate, although it could also involve reconstruction of the topmost SiC layer. After oxidation, the buffer layer structure decouples from the underlying SiC and the √43×√43-R±7.6° pattern disappears. This decoupling behavior is analogous to that which occurs for the 6√3×6√3-R30° buffer layer that forms on the Si-face.[18,19] μ-LEED study of the decoupled C-face buffer layer shows a sixfold arrangement of spots aligned at 30° relative to the primary (1,0) SiC spots, with relatively little rotational disorder, similar to that reported for ML-graphene on the C-face formed by the CCS method.[22]

## Acknowledgements

We gratefully acknowledge support from the National Science Foundation (grant DMR-0856240).

Table I. Sample names, substrate types (n-type 6H-SiC or semi-insulating 4H-SiC), surfaces used, and graphene growth conditions.

| Name | Substrate | Surface | Growth conditions |
|---|---|---|---|
| D1 | n-6H | C-face | $5\times10^{-5}$ Torr disilane, 1220°C, 10 min |
| N1 | s.i.-4H | Si-face | 1 atm neon, 1630 °C, 20 min |
| N2 | s.i.-4H | C-face | 1 atm neon, 1450 °C, 10 min |

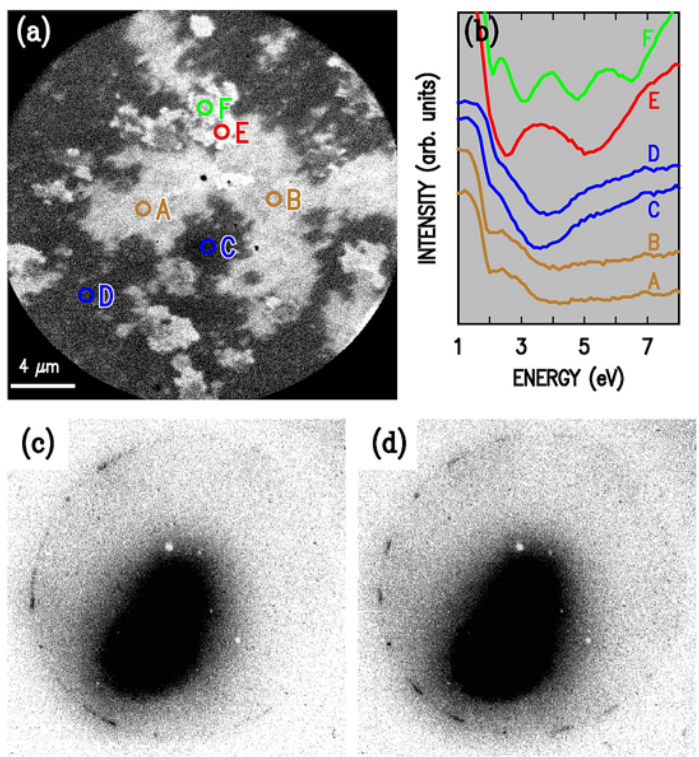

FIG 1. (Color online) Results for graphene on the C-face, sample D1, before oxidation. (a) LEEM image at beam energy of 3.8 eV. (b) Intensity of the reflected electrons from different locations marked in (a). (c) and (d) μ-LEED patterns acquired at 44 eV, using a 5 μm aperture centered at locations A (buffer layer) and C (1-ML graphene on buffer layer) in panel (a), respectively.

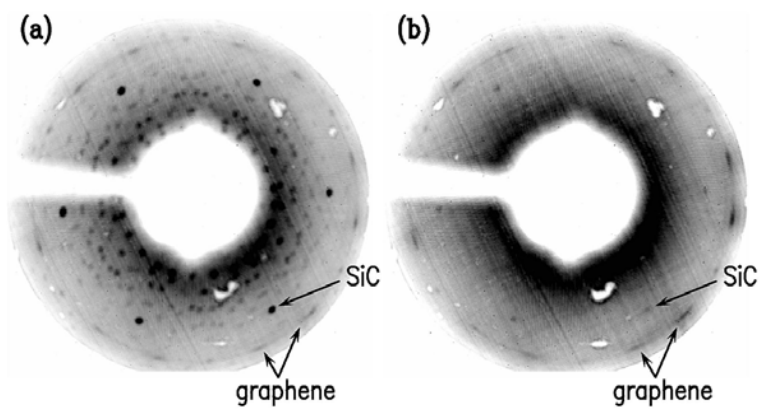

FIG 2. Results for sample D1. (a) and (b) LEED patterns at 100 eV, before and after oxidation of the sample, respectively.

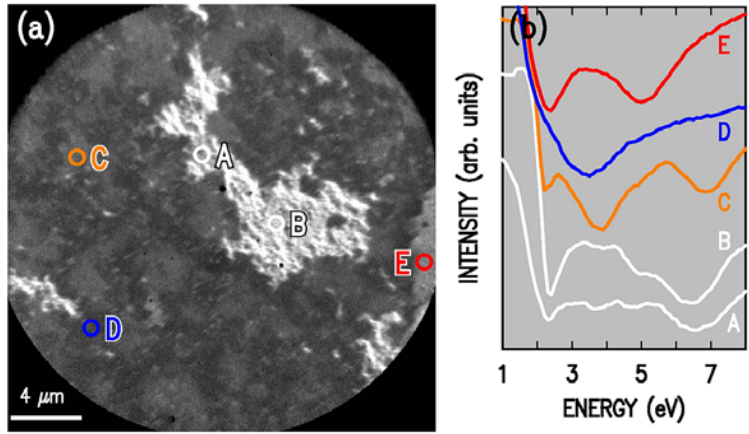

FIG 3. (Color online) Results for graphene on the C-face, sample D1, after oxidation. (a) LEEM image at beam energy of 3.8 eV. (b) Intensity of the reflected electrons from different locations marked in (a).

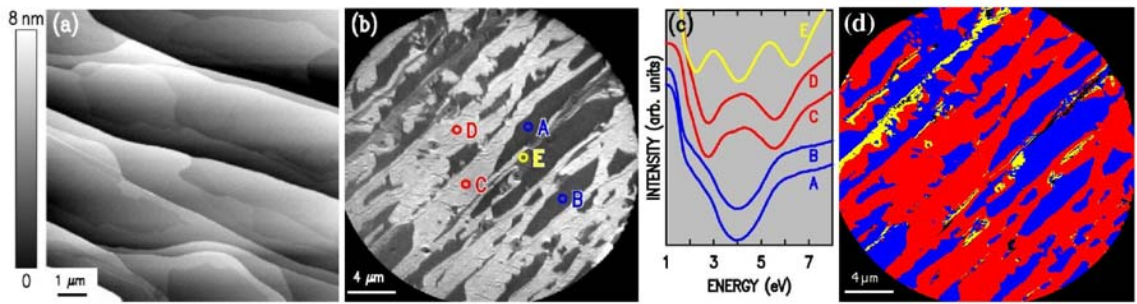

FIG 4. (Color online) Results for graphene on the Si-face, sample N1. (a) AFM image, displayed with gray scale range of 8 nm, (b) LEEM image at beam energy of 3.8 eV, (c) Intensity of the reflected electrons from different locations marked in (b) as a function of electron beam energy, (d) Color-coded map of local graphene thickness; blue, red and yellow correspond to 1, 2, and 3 ML of graphene, respectively.

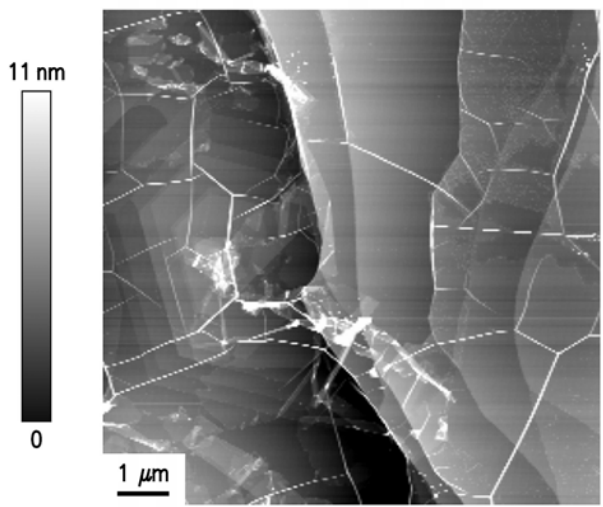

FIG 5. AFM image for graphene on the C-face, sample N2. The image is displayed with gray scale range of 11 nm.

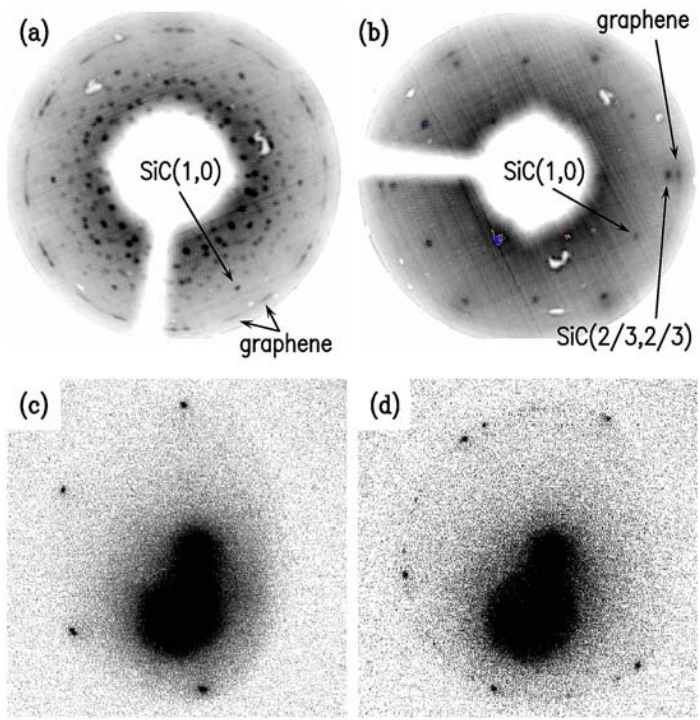

FIG 6. Results for sample N2. (a) and (b) LEED patterns at 100eV, before and after oxidation of the sample, respectively. (c) and (d) μ-LEED patterns acquired after oxidation, with a 5 μm aperture at 44 eV, from the buffer layer and multilayer graphene on the buffer, respectively.